\documentclass[prd,aps,groupedaddress,
preprintnumbers,
nofootinbib]{revtex4}
\usepackage{graphicx}
\usepackage{amsmath}
\usepackage{bm}
\usepackage{epsfig}
\usepackage{amssymb}

\begin{document}
\vspace*{1cm}
\title{Erratum: Order-$\bm{v^4}$ corrections to $\bm{S}$-wave quarkonium decay\\
[0pt][Phys.\ Rev.\ D {\bf 66}, 094011 (2002)]}
\author{Geoffrey~T.~Bodwin and Andrea Petrelli}

\date{March 11, 2013}

\preprint{\begin{tabular}{l}ANL-HEP-PR-12-105\end{tabular}}
\pacs{}

%

\maketitle


In this erratum, we correct several errors in Ref.~\cite{Bodwin:2002hg}. 

In Eqs.~(6.25), (6.26), (6.27), (6.28b), and (6.30) of
Ref.~\cite{Bodwin:2002hg}, the denominator factor $81$ should be written
as $27(D-1)$, where $D$ is the space-time dimension. The factor $D-1$
arises from the projection of bilinears of $p$ onto $S$-wave states.
Here $p$ is one-half the relative momentum of the quark and antiquark in
the quark-antiquark rest frame. Specifically, the $S$-wave projections
yield
\begin{equation}
p_ip_jp_i'p_j'\to 
\frac{\bm{p}^2}{D-1}\delta_{ij}\frac{\bm{p}^2}{D-1}\delta_{ij}
=\frac{\bm{p}^4}{D-1},
\end{equation}
where $\bm{p}^2={\bm{p}'}^2$, but we distinguish the orientations of the
momentum $\bm{p}$ in the incoming state and the momentum $\bm{p}'$ in the
outgoing state in order to project each onto an $S$-wave
angular-momentum state. The denominator factors $27(D-1)$ affect the
$\overline{\rm MS}$ renormalization of the NRQCD matrix element. In
$\overline{\rm MS}$ renormalization, one constructs the counterterm for
a UV-divergent subdiagram by subtracting the poles in $\epsilon=(4-D)/2$
that appear in that subdiagram, plus some associated constants. In order
to ensure that the counterterm contribution removes the contribution
that is proportional to the UV divergence in the divergent subdiagram,
one must compute in $D$ dimensions all factors that are external to the
divergent subdiagram. These external factors include factors that derive 
from the external momenta, such as the projections onto specific
orbital-angular-momentum states. A failure to follow this procedure at
the one-loop level leads to a shift of the result by a constant term.
However, at the level of two or more loops, it leads to a breakdown in
the consistency of the renormalization program, with the consequence
that uncanceled poles appear in short-distance coefficients
\cite{Bodwin:2012xc}. It follows from these corrections that the term
$-833/972$ in Eq.~(6.31c) should be replaced with $-805/972$ and that
the term $7.62$ in the last line of Table~IV should be replaced with
$8.22$.

Equation~(6.3) of Ref.~\cite{Bodwin:2002hg} should read
\begin{equation}
f(p)=\left[\frac{E(p)}{m}\right]^{3(D-2)-D}
=\left[\frac{E(p)}{m}\right]^2\left[1-4\epsilon\log\frac{E(p)}{m}
\right]+O(\epsilon^2).
\end{equation}
This correction does not affect the term of order $\epsilon^0$ in 
Eq.~(6.3). Hence, as is explained just after Eq.~(6.3), it does not 
affect the result of the calculation.

Equation~(6.21) of  Ref.~\cite{Bodwin:2002hg} should read
\begin{eqnarray}
\tilde{\cal M}_{\rm IR}&=&\frac{\bm{p}^4}{m^8}
\frac{(N_c^2-1)(N_c^2-4)}{N_c^2}
\frac{128\pi^3\alpha_s^3}{(3-2\epsilon)^3}\mu^{6\epsilon}
\left\{
\frac{(3-2\epsilon)(1-\epsilon)}{x^2}
-\frac{2(2-\epsilon)y(1-y)}{x^2}\right.
+\frac{(3-2\epsilon)(1-\epsilon)}{(1-xy)^2}\nonumber\\
&&-\frac{2(2-\epsilon)x(1-x)(1-y)}{(1-xy)^4}
\left. +\frac{(3-2\epsilon)(1-\epsilon)}{[1-x(1-y)]^2}
-\frac{2(2-\epsilon)x(1-x)y}{[1-x(1-y)]^4}\right\}.
\end{eqnarray}
The terms that contain factors of $x^2$ in the denominator, the terms
that contain factors of $(1-xy)^2$ in the denominator, and the terms
that contain factors of $[1-x(1-y)]^2$ in the denominator are related to
each other by cyclic permutations of the energy fractions $x_1$, $x_2$,
and $x_3$ and give equal contributions to the integral over the
final-state phase space. This error in Eq.~(6.21) of
Ref.~\cite{Bodwin:2002hg} was not propagated into the remainder of the
calculation.

In Eq.~(2.16) of Ref.~\cite{Bodwin:2002hg}, $0.82n_f$ should be $0.81n_f$. 
An exact expression for $F_{ee}({}^3S_1)$ through order 
$\alpha^2\alpha_s^2$ is \cite{Beneke:1997jm}
\begin{eqnarray}
F_{ee}({}^3S_1)&=&\frac{2\pi Q^2\alpha^2}{3}\bigg\{1
-4C_F\frac{\alpha_s(m)}{\pi}+\bigg[\frac{103}{27}-\frac{511\pi^2}{162}
-\frac{224\pi^2\ln(2)}{27}
-\frac{250\,\zeta(3)}{9}+\frac{22}{27}n_f
+\frac{140\pi^2}{27}\ln\biggl(\frac{2m}{\mu_\Lambda}\biggr)
\biggl(\frac{\alpha_s}{\pi}\biggr)^2\biggr]\biggr\},\nonumber\\
\end{eqnarray}
where $\zeta(z)$ is the Riemann zeta function.

There are several additional errors of a minor nature in
Ref.~\cite{Bodwin:2002hg}. Just after Eq.~(2.10), $m^2 F_1({}^3S_1)$
should be $F_1({}^3S_1)$. Just after Eq.~(2.14), $m^2
F_{\gamma\gamma}({}^1S_0)$ should be $F_{\gamma\gamma}({}^1S_0)$. 
In Eq.~(5.5c), $58/54$ should
be $29/27$. In Eq.~(6.13), $k_1+k_2+k_3$ should be $k_1^0+k_2^0+k_3^0$.

\begin{acknowledgments}

We thank Chaehyun Yu for pointing out the error in Eq.~(6.3) of
Ref.~\cite{Bodwin:2002hg} and Wen-Long Sang for pointing out the error
in Eq.~(6.27) of Ref.~\cite{Bodwin:2002hg}. We also thank Jungil Lee for
bringing to our attention some errors in an earlier version of this
erratum. The work of G.T.B.\ in the High Energy Physics Division
at Argonne National Laboratory was supported by the U.~S.~Department of
Energy, Division of High Energy Physics, under Contract No.\
DE-AC02-06CH11357.
 
\end{acknowledgments}

\end{document}